\documentclass[pra, superscriptaddress, reprint, aps, a4paper, showkeys]{revtex4-1}

\usepackage{graphicx}
\usepackage{amsmath}
\usepackage{amsthm} 
\usepackage{amssymb}
\usepackage{amsbsy}
\usepackage{fullpage}
\usepackage{natbib}
\usepackage[usenames]{color}
\usepackage{caption}
\usepackage{amsbsy}

\newcommand{\ket}[1]{\left|#1\right\rangle}
\newcommand{\bra}[1]{\left\langle #1 \right|}

\newcommand{\op}[1]{\hat{#1}}
\newcommand{\spc}[1]{\mathcal{#1}}

\definecolor{DarkRed}{rgb}{0.7,0,0}

\newcounter{ComntCntr} 

\DeclareMathOperator{\Tr}{Tr}

\begin{document}

\title{Multipartite nonlocality in the presence of particle-number super-selection rules}
\author{Sebastian Meznaric}
\affiliation{Clarendon Laboratory, University of Oxford, Oxford OX1 3PU, United Kingdom}
\author{Libby Heaney}
\affiliation{Clarendon Laboratory, University of Oxford, Oxford OX1 3PU, United Kingdom}
\affiliation{Centre for Quantum Technologies, National University of Singapore, Singapore}
\author{Dieter Jaksch}
\affiliation{Clarendon Laboratory, University of Oxford, Oxford OX1 3PU, United Kingdom}
\affiliation{Centre for Quantum Technologies, National University of Singapore, Singapore}

\begin{abstract}
Super-selection rules severely restrict the possible operations one can perform on an entangled state. Their effect on the observation of non-locality through the Bell inequalities is only partially understood. In this article we examine a range of multipartite entangled states and find that for those states the super-selection rules limit  genuinely multipartite nonlocality but not the non-genuinely multipartite. We argue that when the number of particles is smaller than the number of parties, genuinely multipartite nonlocality cannot be detected unless the super-selection rules can be relaxed. The tests proposed in the paper can be implemented using routinely used experimental techniques. 
\end{abstract}

\maketitle

\section{Introduction}

Quantum theory exhibits the strange, and to some spooky, phenomenon of nonlocality. Nonlocality refers to the statistical results of local measurements on a composite system that cannot be explained by a local hidden variable theory. A local hidden variable theory is any theory where objects cannot exchange information faster than at the speed of light and where all measurable properties of an object are well defined prior to the measurement. 

Bell and others \cite{Bell64, Clauser69} showed that any local hidden variable theory must satisfy an inequality known as Bell's inequality. For these inequalities to be violated the underlying quantum state must be entangled. While entanglement of distinguishable particles is well understood, the situation becomes less clear when the particles are indistinguishable \cite{Simon02, Amico08} due to the (anti)-symmetrization. In this case it is not immediately clear what the canonical tensor product structure should be. One possible choice is to factorise the Hilbert space according to where the particles are detected. Second quantized spatial field modes are used for this purpose \cite{Peres95}. 

Just like ordinary states, the spatial field modes could exhibit entanglement in the particle number, or Fock, basis \cite{VanEnk05}. As an example, consider the Bell state formed from a single particle distributed over two spatial field modes $a$ and $b$,
\begin{align}
	\ket{\psi} = \frac{1}{\sqrt{2}} \left( \ket{1_a}\ket{0_b} + \ket{0_a}\ket{1_b} \right). \label{eq:BellState}
\end{align}
Here the state $\ket{1_a}$ indicates the presence of a particle in the mode $a$, while the state $\ket{0_a}$ indicates the absence of any particles in that mode. The state is not separable in this basis and could thus manifest nonlocal features should a Bell inequality test be attempted. However, in order to perform a Bell test one would need to measure the spatial field modes in more than one basis. While one basis could be the particle number basis, the other would be a basis formed from superpositions of particle number states.

This is not as simple as it may appear. When dealing with massive particles, observation of coherent particle number superpositions, such as $\frac{1}{\sqrt{2}}\left(\ket{0_a} + \ket{1_a}\right)$, is no longer possible \cite{Wick52}, restricting our ability to conduct a Bell inequality test. This is known as a super-selection rule. Inability to measure such states would seemingly also restrict the possible utilization of entanglement as a resource in quantum communication \cite{Heaney09-1}. 

Since directly measuring the superpositions is futile, research has been conducted on alternative ways to effectively produce the same results as if it were possible to measure superpositions directly. The presence of an additional, jointly prepared but separable, state can \emph{activate} the nonlocality in mode entangled states \cite{White08, Paterek11}. These additional states $\rho_{RF}$ are usually referred to as the \emph{reference frames}. So while a mode-entangled state $\rho$ may not be able to violate Bell inequalities on its own, the state $\rho \otimes \rho_{RF}$ can do so. This is the result of the extended Hilbert space and the ability of the reference frame to track the phase relationship between the modes. In \cite{Paterek11} they also show that locally prepared reference frames are not useful for nonlocality activation since there is no phase relationship between the subsystems. 

A wealth of other literature exists on the subject dealing with the bipartite case. Early work by Aharonov and Vaidman \cite{Aharonov00} considers whether the phase information can be extracted from a single particle wave and discusses the difference between fermions and bosons. Another early work on the topic includes the consideration of entanglement in Bose-Einstein condensates by Simon \cite{Simon02}. Bartlett, Rudolph and Spekkens give a good overview of reference frames and super-selection rules (SSR) in \cite{Bartlett07}. In \cite{Bartlett06} the same authors consider the wider debate over the existence of quantum coherent superpositions between eigenstates of certain quantities. In \cite{Heaney09-1} Heaney and Vedral show how mode entangled states subject to super-selection rules can be used as a resource for quantum communication by relaxing the super-selection rules through the use of particle reservoirs. Related to our present work, Ashhab, Maruyama and others consider the possibility of nonlocality violation via Bell inequalities in mode entangled states in \cite{Ashhab07, Ashhab07-1, Ashhab09}.

While the bipartite nonlocality of modes is fairly well understood,  multipartite nonlocality of modes under super-selection rules has not yet been investigated. Multipartite Bell inequalities allow to test for this type of nonlocality. Two different types of nonlocality can be observed in the multipartite setting - genuinely and nongenuinely multipartite. The issue here is that several components of the system could form a nonlocal part of the whole while the rest of the system remains local. A genuinely multipartite nonlocal system cannot be broken down into local and nonlocal parts - it is nonlocal as a whole. In contrast such a breakdown is possible for nongenuinely multipartite nonlocal systems. The Svetlichny inequality \cite{Svetlichny87} can be used to test for genuinely 3-partite nonlocality and  it has been generalized by Bancal et al. in \cite{Bancal11} to multiple parties. On the other hand, nongenuinely multipartite inequalities have been fully characterised by \.{Z}ukowski and Brukner \cite{Zukowski02} and Werner and Wolf \cite{Werner01}. Multipartite Bell-type inequalities also exist for the purpose of detection of entangled states (entanglement witness), as seen in \cite{Durkin05}.

The inherent difference between genuinely and non-genuinely multipartite nonlocality leads us to ask whether the super-selection rules restrict the type of nonlocality we can observe. Since genuinely multipartite nonlocality is the most powerful type of nonlocality, the inability to observe it would imply that the super-selection rules restricted mode-entangled states are inherently less powerful than the more conventional entangled states where each party is given a particle with multiple internal states.


To study this issue we shall test a selection of mode entangled states, including W-states and Dicke states, for multipartite nonlocality. Our approach, following \cite{Heaney10}, is to use multiple copies of the state $\rho$ in place of a reference frame, i.e. $\rho \mapsto \bigotimes_n \rho$. The benefit of this is that one simply generates $n$ copies of the state and measures the modes using atomic beamsplitters \cite{Moura04, Palmer05, Bongs04} . However, while using pairs of states does indeed activate nongenuinely multipartite nonlocality it does not activate genuinely multipartite nonlocality. Without super-selection rules both nongenuinely and genuinely multipartite nonlocality is present. 

In the discussion we shall argue a necessary condition for observation of genuinely multipartite nonlocality in the presence of super-selection rules is that the number of particles in the state is greater than or equal to the number of parties. The argument applies only when no reference frame or a particle reservoir exists that could be used to lift or relax the super-selection rules. Notice that this is only a necessary but not sufficient condition. States with three particles distributed among three parties that do not violate any kind of nonlocality trivially exist. 

Understanding of nonlocality is also important for the purposes of deciding how difficult a particular state would be to simulate classically. The states exhibiting nonlocality can be difficult to simulate \cite{Horodecki09}. It is expected that genuinely multipartite nonlocality would be harder to simulate than the non-genuinely multipartite. Similarly to simulating the behaviour of such states, the statistics of measurement results observed with such states is also difficult to simulate classically. It was shown in \cite{Massar01} that simulating local measurement results of $n$ Bell states classically would require $O(2^n)$ bits of classical communication. In addition to a large amount of communication required, in the classical case the measurement results are not available instantly as in the quantum case.

The paper is organised as follows.  Section \ref{sec:BellIntro} contains the introduction to the multipartite Bell-type inequalities for mode entangled states in the presence of superselection rules. Section \ref{subsec:Genuine} will then apply the techniques from section \ref{sec:BellIntro} to genuinely multipartite Bell inequalities followed by the non-genuinely multipartite Bell inequalities in section \ref{subsec:Nongenuine}. We shall then apply derived genuinely and non-genuinely Bell inequalities to a selection of states in section \ref{sec:Applications} and conclude with a discussion and conclusion in sections \ref{sec:Discussion} and \ref{sec:Conclusion}, respectively. 

\section{Multipartite Bell-type inequalities in the presence of SSR} \label{sec:BellIntro}

For our purposes we shall think of the Bell-type inequalities as any inequality of the type
\begin{align}
	\left|f(\op{\alpha}_1, \op{\beta}_1, \op{\alpha}_2, \op{\beta}_2 \ldots, \op{\alpha}_M, \op{\beta}_M)\right| \leq B, \label{eq:GeneralBell}
\end{align}
where $\op{\alpha}_k$ and $\op{\beta}_k$ are the measurement settings chosen by the local observers, $B$ is the bound and $f$ is a function of the correlation functions. While quantum mechanically the measurement settings $\op{\alpha}_k$ and $\op{\beta}_k$ can be replaced with observables, it should be noted that Bell-type inequalities can be applied to any theory. We also require that $f$ be such that the value of LHS larger than $B$ implies no local hidden variable theory can reproduce the statistics leading to that value. We shall assume that the local measurement devices have two possible measurement outcomes, $+1$ and $-1$. When the average of the left hand side exceeds the value on the right hand side we say that the Bell inequality has been violated. In the presence of a reference frame state $\rho_{RF}$ in addition to the observed state $\rho$ the observables $\op{\alpha}_k$ and $\op{\beta}_k$ act on the appropriate local subsystems of both $\rho_{RF}$ and $\rho$. 

Quantum mechanically, the measurement settings $\op{\alpha}_k$ and $\op{\beta}_k$ are quantum observables with eigenvalues $+1$ and $-1$. Similarly the correlation functions can be written simply as the expectation values of the tensor product of the observables, i.e.
\begin{align}
	E(\alpha_1, \ldots \alpha_M) = \left\langle \alpha_1 \otimes \ldots \otimes \alpha_M \right \rangle.
\end{align} 
But while in the absence of super-selection rules one could choose any observables in place of $\op{\alpha}_i$ and $\op{\beta}_i$, observables we choose must be expressible as linear combinations of operators projecting on SSR-compatible states (i.e. states containing no superpositions of different mass). Therefore for massive particles all observables satisfying super-selection rules are required to commute with the particle-number operator $\op{N}$.

\begin{figure*}[hbt]
	\centering
	\includegraphics[scale=1.00]{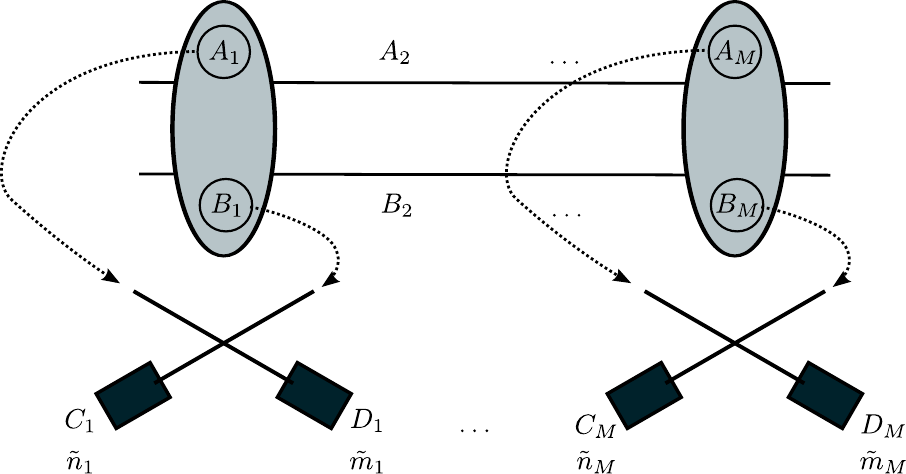}
	\caption{A graphical representation of modes of the two states and the outputs from a beamsplitter.}
	\label{fig:ModesDiagram}
\end{figure*}

The state in eq. (\ref{eq:BellState}) can in general behave just like any other maximally entangled state, as long as the particle is massless (see for instance \cite{Babichev04, O'Brien09}). Our system is split into $M$ spatially separated modes, with the total number of massive particles in the system being $N$ (see figure \ref{fig:ModesDiagram}).  Our Hilbert space $\spc{H}$ must therefore effectively decompose as
\begin{align}
	\spc{H} = \bigoplus_{n_1 + \ldots + n_M = N} \left( \spc{H}^1_{n_1} \otimes \spc{H}^2_{n_2} \otimes \ldots \otimes \spc{H}^M_{n_M} \right),
\end{align}
where the direct sum is over all the partitions of $N$ into $n_1 + n_2 + \ldots + n_M$ with $n_k$ being non-negative integers. This restriction leaves our states with less operationally meaningful entanglement than may otherwise be possible \cite{Meznaric11} and therefore may also restrict the ability to observe Bell inequality violations. 

However, in \cite{Aharonov67} Aharonov and Susskind show that observing superpositions \emph{is} possible in the presence of additional reference frame states. The ability to locally observe superpositions is crucial in order to observe the Bell inequality violations. It was shown in \cite{Heaney10} that it is possible to partially overcome the restrictions on measurements by using two or more copies of the same state, since Alice and Bob can now locally conduct measurements on both of the states simultaneously (see figure \ref{fig:ModesDiagram}).

Since we have two or more copies of our state we shall denote the modes corresponding to the first copy as $A_k$ and those to the second one as $B_k$. The index $k$ refers to the local observer. The Fock basis at each of the modes shall be labelled as $\ket{n_k}$ for modes $A_k$ and $\ket{m_k}$ for modes $B_k$, while the operators $\op{a}_k$ and $\op{b}_k$ are the annihilation operators for modes $A_k$ and $B_k$, respectively. 

Each of the observers has a beamsplitter whose inputs are connected to $A_k$ and $B_k$ modes, as shown in figure \ref{fig:ModesDiagram}. The outputs are labelled as $C_k$ and $D_k$, with the Fock basis on these modes being labelled by $\ket{\tilde{n}_k}$ and $\ket{\tilde{m}_k}$ respectively. The annihilation operators of input and output modes are related by
\begin{align}
	\op{c}_k & = \cos(\theta_k) \op{a}_k + \sin(\theta_k) e^{-i \phi_k} \op{b}_k \\
	\op{d}_k & = \sin(\theta_k) \op{a}_k - \cos(\theta_k) e^{-i \phi_k} \op{b}_k.
\end{align}
The state $\ket{\tilde{n}_k, \tilde{m}_k}$, where the integers $\tilde{n}_k$ and $\tilde{m}_k$ denote the number of particles in each of the beamsplitter output modes respectively, can be written as
\begin{align}
	\ket{\tilde{n}_k, \tilde{m}_k} = \frac{\left(\op{c}_k^\dagger\right)^{n_k}}{\sqrt{n_k!}} \times \frac{\left(\op{d}_k^\dagger\right)^{m_k}}{\sqrt{m_k!}},
\end{align}
where $\ket{0_k, 0_k}$ is the vacuum state, $\theta_k$ is the mixing parameter between the two modes for a single observer and $\phi_k$ is the phase relationship between the two modes. The list of possible output states for two copies of single-particle input states can be seen in table  \ref{tab:Measurements}. The states with tilde refer to the output modes of the beamsplitter while those without are the original modes. These parameters are properties of the beamsplitter and are used to switch between the different measurement settings for each observer. Note that this can be easily generalised to multiple copies of the state, replacing the beamsplitters with multiport devices as in \cite{Vourdas05}. Experimental implementation of multiport devices is considered in \cite{Peruzzo11}.

The observers at each of the local pairs of modes can measure one of the observables specified by
\begin{align}
	\op{O}_k\left(\phi_k, \theta_k \right) = \sum_{\tilde{n}_k + \tilde{m}_k = 0}^{\tilde{n}_k + \tilde{m}_k = N} \epsilon(\tilde{n}_k, \tilde{m}_k) \ket{\tilde{n}_k, \tilde{m}_k}\bra{\tilde{n}_k, \tilde{m}_k}.
\end{align}
The function $\epsilon(\tilde{n}_k, \tilde{m}_k)$ is the binning function, deciding whether the measurement outcome should correspond to $+1$ or to $-1$.  We follow the approach in \cite{Heaney10} and choose the binning function to be
\begin{align}
	\epsilon(\tilde{n}_k, \tilde{m}_k) = (-1)^{\tilde{m}_k + \frac{1}{2}\left(\tilde{m}_k+\tilde{n}_k\right)\left(\tilde{m}_k+\tilde{n}_k+1\right)}.
\end{align}
A sharp binning function that alternates every time the outcome changes by 1 was shown in \cite{Lee09} to lead to a tight and optimal bipartite Bell inequality. 

Consider the case of having two copies of a single-particle state (table \ref{tab:Measurements}). Notice that for some values of $\theta$ some outcomes have vanishing probability of being measured. For instance, the output state $\ket{\tilde{1}_k, \tilde{1}_k}$, for the measurement setting $\theta_k=\pi/4, 3\pi/4$ etc., will result from superpositions of two particles in a single input mode. This contradicts our initial assumption of single particle input states. Therefore, while for the stated value of $\theta$ the state $\ket{\tilde{1}_k, \tilde{1}_k}$ is non-vanishing, it's probability of detection vanishes. This simplifies the calculation somewhat. A similar phenomenon occurs for the states $\ket{\tilde{2}_k, \tilde{0}_k}$ and $\ket{\tilde{0}_k, \tilde{2}_k}$ when $\theta_k = 0, \pi/2, \pi$ etc.
\begin{center}
	\begin{table*}[ht]
	{\small
	\hfill{}
	\begin{tabular}{ccc}
	$\ket{\tilde{n}_k, \tilde{m}_k}$ & Measurement in $\ket{n_k, m_k}$ basis & $\epsilon(\tilde{n}_k, \tilde{m}_k)$ \\ \hline
	$\ket{0_k, 0_k}$ & $\ket{\tilde{0}_k, \tilde{0}_k}$ & 1 \\
	$\ket{\tilde{1}_k, \tilde{0}_k}$ & $\cos(\theta_k) \ket{1_k, 0_k} + \sin(\theta_k) e^{i \phi_k} \ket{0_k, 1_k}$ & -1 \\
	$\ket{\tilde{0}_k, \tilde{1}_k}$ & $\sin(\theta_k) \ket{1_k, 0_k} - \cos(\theta_k) e^{i \phi_k} \ket{0_k, 1_k}$ & 1 \\
	$\ket{\tilde{2}_k, \tilde{0}_k}$ & $\cos^2(\theta_k) \ket{2_k, 0_k} + \sqrt{2} \cos(\theta_k) \sin(\theta_k) e^{i \phi_k} \ket{1_k, 1_k} + e^{2i \phi_k} \sin^2(\theta_k) \ket{0_k, 2_k}$ & -1 \\
	$\ket{\tilde{1}_k, \tilde{1}_k}$ & $\sqrt{2} \cos(\theta_k) \sin(\theta_k) \left( \ket{2_k, 0_k} - \ket{0_k, 2_k}\right) - \cos(2\theta_k) e^{i \phi_k} \ket{1_k, 1_k}$ & 1 \\
	$\ket{\tilde{0}_k, \tilde{2}_k}$ & $\sin^2(\theta_k) \ket{2_k, 0_k} - \sqrt{2} \cos(\theta_k) \sin(\theta_k) e^{i \phi_k} \ket{1_k, 1_k} + \cos^2(\theta_k) e^{2i\phi_k} \ket{0_k, 2_k}$ & -1
	\end{tabular}}
	\hfill{}
	\caption{The table shows the local measurements on the single-particle state after the action of the beamsplitter. \label{tab:Measurements}} 
	\end{table*}
\end{center}
For our parametrisation of measurements the correlation function can be written as
\begin{align}
	E(\phi_1, \theta_1, \phi_2, \theta_2, \ldots \phi_M, \theta_M) = \Tr\left[ \rho \bigotimes_k \op{O}_k(\phi_k, \theta_k)\right].
\end{align}
Similarly the multipartite Bell inequality can be obtained by assigning $\op{\alpha}_k = \op{O}(\phi_k^A, \theta_k^A)$ and $\op{\beta}_k = \op{O}(\phi_k^B, \theta_k^B)$ in the general Bell-type inequality in eq. (\ref{eq:GeneralBell}).
In the next section we shall use our operators $\op{O}_k$ with specific genuinely and nongenuinely multipartite Bell inequalities. 

\subsection{Genuinely multipartite Bell inequalities}\label{subsec:Genuine}

We shall first consider genuinely 3-partite nonlocality and then expand the concept to arbitrary number of parties. Genuinely 3-partite nonlocality occurs when the correlations are not limited to multiple 2-partite correlations. To make the concept more precise, consider $p(a,b,c)$ to be the probability of the respective parties obtaining the results $a$, $b$ and $c$ (Alice obtains $a$, Bob obtains $b$ and Charlie obtains $c$). Genuinely multipartite nonlocality is obtained whenever there is no hidden-variable model that could factor the probabilities $p(a,b,c)$ as
\begin{align}
	p(a,b,c) = \int q(a,b | \lambda) r(c | \lambda) d\mu(\lambda), \label{eq:GenuineDist}
\end{align}
where $\lambda$ is the hidden variable, $d\mu$ is the probability measure and $q$ and $r$ are the conditional probabilities of obtaining the specified outcomes given the value of the hidden variable. Contrast this with the less restrictive non-genuinely 3-partite nonlocality that only assumes that the probabilities cannot be written as
\begin{align}
	p(a,b,c) = \int q(a | \lambda) r(b | \lambda) s (c | \lambda) d\mu(\lambda), \label{eq:NongenuineDist}
\end{align}
where $q$, $r$ and $s$ are again the conditional probabilities. 
While the eq. (\ref{eq:GenuineDist}) allows for two-partite correlations (see next subsection) the distribution in eq. (\ref{eq:NongenuineDist}) does not. It is in this sense that we refer to the 3-partite distributions that cannot be factored as in eq. (\ref{eq:GenuineDist}) as genuinely multipartite nonlocal, while those that merely cannot be factored as in eq. (\ref{eq:NongenuineDist}) as nongenuinely multipartite nonlocal.

In \cite{Bancal11}, Bancal \textit{et al.} extended the Svetlichny inequality to the general multipartite case. They obtained a sequence of $n$-partite Bell inequalities $\mathcal{B}_n$. Instead of assuming that the probability distribution is not of the form of eq. (\ref{eq:GenuineDist}), their approach assumes the probability distribution is not of the form
\begin{align}
	p(a_1, a_2, \ldots, a_n) = \int p(a_1, \ldots, a_{n-1} | \lambda) p(a_n | \lambda) d\mu(\lambda). \label{eq:GenuineMultipartiteDist}
\end{align}
Assuming such a distribution they show that the following inequality must then be satisfied by any system that is not genuinely multipartite nonlocal
\begin{align}
	\langle \op{\mathcal{B}}_n \rangle =  \langle \op{\alpha}_n \op{\mathcal{B}}_{n-1} + \op{\beta}_n \op{\mathcal{B}}'_{n-1} \rangle \leq 2^{n-1}, \label{eq:MultipartiteBell}
\end{align}
where $\op{\alpha}_n$ and $\op{\beta}_n$ are the two measurement settings to be performed by the $n$-th party. $\op{\mathcal{B}}_n$ is the Bell inequality operator for $n$ parties and $\op{\mathcal{B}}'_n$ is the same as $\op{\mathcal{B}}_n$ but with the measurement settings $\op{a}$ and $\op{b}$ reversed. The sequence starts at $n=2$ with $\op{\mathcal{B}}_2$ being the CHSH inequality. The Svetlichny inequality is obtained when $n=3$. It must then be the case that whenever inequality (\ref{eq:MultipartiteBell}) is violated the probability distribution of the underlying physical system is not of the form (\ref{eq:GenuineMultipartiteDist}). We can therefore conclude that our system possesses genuinely multipartite nonlocality. 

The above inequality can easily be applied to the mode-entangled system of two state copies by replacing $\op{\alpha}_k$ and $\op{\beta}_k$ with $\op{O}(\phi_k^A, \theta_k^A)$ and $\op{O}(\phi_k^B, \theta_k^B)$, respectively. 

\pagebreak

\subsection{Non-genuinely multipartite Bell inequalities}\label{subsec:Nongenuine}

In contrast to genuinely multipartite nonlocal probability distributions, the non-genuinely multipartite nonlocal disitributions are those that cannot be written as
\begin{align}
	p(a_1, \ldots a_k) = \int q_1(a_1 | \lambda) q_2(a_2 | \lambda) \ldots q_n(a_n | \lambda) d\mu(\lambda). \label{eq:NongenuineBell}
\end{align}
Nongenuinely multipartite nonlocal distributions thus still posses nonlocality in the sense that the probability distribution cannot be factored as in eq. (\ref{eq:NongenuineBell}) but it may be possible to split the subsystems among two groups and factor the probability distribution with respect to those as in (\ref{eq:GenuineMultipartiteDist}). The word non-genuinely therefore does not refer to the nonlocality, it refers to the multipartiteness of the nonlocality. 

Let us write our operators $\op{\alpha}_k$ and $\op{\beta}_k$ as $\op{\alpha}_k = \op{q}_k(1)$ and $\op{\beta}_k = \op{q}_k(2)$. The \.{Z}ukovski-Brukner inequality \cite{Zukowski02} can be written as
\begin{widetext}
\begin{align}
	\Biggl\langle\sum_{s_1, \ldots, s_M = -1,1} \biggl| \sum_{k_1, \ldots, k_M = 1,2} s_1^{k_1-1} s_2^{k_2-1} \ldots s_M^{k_M-1} \op{q}_1(k_1) \otimes \op{q}_2(k_2) \otimes \ldots \otimes \op{q}_M(k_M) \biggr|\Biggr\rangle \leq 2^M. \label{eq:NongenuineGeneralBell}
\end{align}
For the case of two particles it can be checked by a direct calculation that the inequality reduces to
\begin{align}
	\Bigl\langle\bigl(\bigl|\op{\alpha}_1-\op{\beta}_1\bigr|+\bigl|\op{\alpha}_1+\op{\beta}_1\bigr|\bigr)
   \bigl(\bigl|\op{\alpha}_2-\op{\beta}_2\bigr|+\bigl|\op{\alpha}_2+\op{\beta}_2\bigr|\bigr)\Bigr\rangle \leq 4,
\end{align}
which is the CHSH inequality multiplied by two. Whenever the inequality
\begin{align}
	\Biggl\langle\biggl| \sum_{s_1, \ldots, s_M = -1, 1} S(s_1, \ldots s_M) \sum_{k_1, \ldots, k_M = 1,2} s_1^{k_1 - 1} \ldots s_M^{k_M - 1} \op{q}_1(k_1) \otimes \op{q}_2(k_2) \ldots \otimes \op{q}_M(k_M) \biggr| \Biggr\rangle \leq 2^M \label{eq:NongenuineLessG},
\end{align}
\end{widetext}
is violated, the inequality in eq. (\ref{eq:NongenuineGeneralBell}) is also violated (the converse is not true). Here the function $S$ can take value $-1$ or $+1$ and there are $2^{2^M}$ functions of this form. This form of the inequality is easier to check than the more general form in (\ref{eq:NongenuineGeneralBell}). As they further show in \cite{Zukowski02}, the eq. (\ref{eq:NongenuineLessG}) reduces to the Mermin-Ardehali-Belinskii-Klyshko (MABK) inequalities \cite{Mermin90, Ardehali92, Belinsky93} when the function $S$ is chosen to be
\begin{multline}
	S(s_1, s_2, \ldots, s_M) = \\ \sqrt{2} \cos\left[\frac{\pi}{4}\left(s_1 + \ldots + s_M - M - 1\right)\right].
\end{multline}
For $M = 3$ this reduces to the Bell inequality of the form
\begin{align}
	\Bigl| \hspace{0.5mm} \bigl\langle \op{\alpha}_1 \op{\beta}_2 \op{\beta}_3 + \op{\beta}_1 \op{\alpha}_2 \op{\beta}_3 + \op{\beta}_1 \op{\beta}_2 \op{\alpha}_3 - \op{\alpha}_1 \op{\alpha}_2 \op{\alpha}_3\bigr\rangle \Bigr| \leq 2.
\end{align}
Compared to the Svetlichny inequality, eq. (\ref{eq:MultipartiteBell}) with $n=3$, with eight distinct terms, the above inequality only contains four distinct terms. The additional terms in Svetlichny inequality allow to test for the additional nonlocality.

Similarly to the genuinely multipartite inequalities, we can adapt inequality (\ref{eq:NongenuineLessG}) into inequalities that can be applied to states governed by superselection rules by replacing the operators $\op{\alpha}_k$ and $\op{\beta}_k$ with $\op{O}(\phi_k^A, \theta_k^A)$ and $\op{O}(\phi_k^B, \theta_k^B)$. In the next section we shall use the inequalities derived herein to examine the multipartite nonlocality properties of several mode-entangled states. 

\section{Applying the Inequalities to multipartite mode-entangled states}\label{sec:Applications}

In the following sections we shall compute the Bell inequality violations for W-states and Dicke states. We tested the Bell inequalities for a range of other states as well but have found the results to be either the same as for the selected states or without any violations at all. 

The states we tested invariably display nongenuinely multipartite nonlocality while no genuinely multipartite nonlocality was detected with our method. General reasons for this disparity are given in the discussion section \ref{sec:Discussion}.

\subsection{W-state}

The $M$-party W-state is the state
\begin{multline}
	\ket{W} = \frac{1}{\sqrt{M}} \left( \ket{100 \ldots 0} + \ket{0100 \ldots 0} \right. \\
	\left. + \ldots + \ket{00\ldots 01}\right).
\end{multline}
When working in the particle-number basis the state becomes
\begin{align}
	\ket{W} = \frac{1}{\sqrt{M}}\sum_{k=1}^M \op{a}_k^\dagger \ket{0},
\end{align}
where $\op{a}_k^\dagger$ is a creation operator for $k$-th mode and $\ket{0}$ is a vacuum state. This state can be created by putting a single particle into a delocalised ground state of a balanced optical lattice.

We first tested the W-state without super-selection rules where we only found violations of genuinely multipartite Bell inequality when $M=3$. Therefore we only check genuinely multipartite inequalities (\ref{eq:MultipartiteBell}) for multiple copies of the W-state subject to super-selection rules for $M=3$. We first conducted this test with two copies of the state. When $M=2$ there is no distinction between genuine and non-genuine multipartiteness, but the Bell inequality is clearly violated for $M=2$ since the W-state reduces to one of the Bell states.  We numerically optimised the operators to get the maximum violation, but found that the inequalities were always satisfied. Therefore our scheme detected no genuinely multipartite nonlocality. Note that for a similar scheme \cite{Heaney10}, a violation was found of the CHSH inequality for pairs of superselection rule restricted states. 

However, nongenuinely multipartite inequalities give a different picture. We tested the MABK inequalities on the W-states of up to 5 parties, similarly as genuinely multipartite inequalities. We detected the violation for every $M \geq 2$ (see table \ref{tab:Wstatetable}). In the absence of super-selection rules the W-state violates nongenuinely multipartite Bell inequalities for all $N$.

\begin{center}
	\begin{tabular}{rrrr}
		$M$ &  Nongenuine & Bound & Percentage above\\ \hline \hline
		2 & 2.41421 & 2 & 20.7\% \\
		3 & 4.29929 & 4 & 7.48\% \\
		4 & 8.32456 & 8 & 4.06\% \\
		5 & 16.3915 & 16 & 2.45\%
	\end{tabular}
	\captionof{table}{The table lists the values we obtained numerically for nongenuinely multipartite Bell inequalities with two copies of $M$-party W-states, the bound for $M$-partite inequalities and the percentage by which that bound was exceeded. We include the case $M = 2$ for reference. \label{tab:Wstatetable}}
\end{center}

The same calculation was conducted with the W-state with two particles instead of one, i.e. 
\begin{align}
	\frac{1}{\sqrt{3}}\left(\ket{200} + \ket{020} + \ket{002}\right),
\end{align}
and identical results were obtained for both genuinely and non-genuinely multipartite inequalities. This is not surprising as these states can be obtained from single particle equivalents simply by mapping $\ket{1} \mapsto \ket{2}$. But while the maximum violation is identical, the angles $\phi_k$ in the correlation function become $2\phi_k$. Higher phase sensitivity of these states is to be expected as they are widely used for phase detection (see for example \cite{Durkin07}).

We also checked for what happens to multiple copies of these states and found similar results - only non-genuinely multipartite nonlocality is detected. 

\subsection{Dicke states}

In second quantization Dicke states can be written as
\begin{align}
	\ket{D_{M,N}} = \binom{M}{N} \sum_{\op{P} \in S_N} \op{P}\bigl[ \bigotimes_{k=1}^{N} \op{a}_k^\dagger \ket{0}\bigr],
\end{align}
where $S_N$ is the group of permutation of $N$ elements, $\op{P}$ are the faithful representation matrices of the group on the Fock space, $\ket{0}$ is the vacuum state, $\op{a}_k^\dagger$ is the creation operator. As previously, $M$ represents the number of modes while $N$ represents the number of particles. For $N = 1$ the Dicke states reduce to W-states. Dicke states were originally considered in \cite{Dicke54} as eigenstates of the Hamiltonian describing a gas of bosonic particles with two internal states confined inside a small box. Fermionic states are described similarly with the symmetrization operator replaced with the anti-symmetrizer. 

An interesting set of Dicke states are those with about half the modes containing a particle and the other modes being in the vacuum state $\ket{D_{N, \lfloor N/2 \rfloor}}$. Here $\lfloor x \rfloor$ is the floor function, giving the greatest integer $n$ such that $n \leq x$. They are interesting because they are the Dicke states with the greatest violation of the Bancal Bell inequalities (we tested this numerically up to $N = 6$). In fact, for $N=4,5$ we found them to be the only Dicke states with any Bell inequality violation. We therefore tested the $\ket{D_{N, \lfloor N/2 \rfloor}}$ states for Bell inequality violations in the presence of super-selection rules. We found very similar results to those with W-states (see table \ref{tab:DickestateTable}). As before, only nongenuinely multipartite inequalities were violated.

\begin{center}
	\begin{tabular}{rrrr}
		$M$ & Genuine & Nongenuine & Bound\\ \hline \hline
		3 & 4 & 4.29929 & 4\\
		4 & 4.64821 & 8.38189 & 8\\
		5 & 8.32 & 16.5305 & 16
	\end{tabular}
	\captionof{table}{The table lists the Bell inequality violations for the Dicke states, for both genuinely and nongenuinely multipartite inequalities. \label{tab:DickestateTable}}
\end{center}

\section{Discussion} \label{sec:Discussion}

We applied genuinely and nongenuinely multipartite Bell inequalities to a selection of mode-entangled states in the previous section. Invariably we found that we could not detect any genuinely multipartite nonlocality, while nongenuinely multipartite nonlocality was present. In the introduction we asked the question of whether it is the mode-entangled states that simply do not contain any genuinely multipartite nonlocality or is the present method insufficient to detect it. We also hinted at a relationship between the number of particles in mode entangled state and the nature of nonlocality one can detect. In the following paragraphs we shall explain the issue in more detail. 

First notice that \emph{any} state may be written in second-quantized form and thus any entangled state could be considered to be mode-entangled. A Bell state correlation of two particles' spin states, for instance, can be written in the mode-entangled form via the dual-rail encoding whereby $\ket{0} \mapsto \op{a}^\dagger \ket{0_a, 0_b}$ and $\ket{1} \mapsto \op{b}^\dagger \ket{0_a, 0_b}$. The $\ket{\phi^+}$ Bell state would then be $\frac{1}{\sqrt{2}}\left(\ket{01} \otimes \ket{01} + \ket{10} \otimes \ket{10}\right)$. This state can always be used to violate the Bell inequalities maximally since any combination of $\ket{0_a, 1_b}$ and $\ket{1_a, 0_b}$ can be measured in the presence of super-selection rules. Thus the class of mode entangled states contains maximally nonlocal states.


Next notice that there is a relationship between the number of particles and effective entanglement. Namely, when super-selection rules are in place no genuinely multipartite nonlocality can be observed when the number of particles $N$ is smaller than the number of local observers $M$ - provided the SSR have not been relaxed through the use of reference frames or other means. To see this, first note that Wiseman and Vaccaro \cite{Wiseman03} proposed in 2003 that the effective state in the presence of super-selection rules and local observers, $\tilde{\rho}$, can be obtained from $\rho$ by
\begin{align}
	\tilde{\rho} = \sum_{n_1 + \ldots + n_M = N} \Pi_{n_1, \ldots, n_M} \rho \Pi_{n_1, \ldots, n_M}, \label{eq:Wiseman}
\end{align}
where $\Pi_{n_1, \ldots, n_M} = \bigotimes_{j=1}^M \ket{n_j}\bra{n_j}$. When SSRs are in place the state $\tilde{\rho}$ is equally powerful for quantum communication protocols as $\rho$ (see \cite{Meznaric11}). The sum here is over all partitions of integer $N$ (number of particles) into sums of $M$ integers (number of local observers). The states $\ket{n_j}$ represent $n_j$ particles in any of the modes belonging to observer $j$. Now consider our $N$-particle $M$-party state. It can be written as a superposition of different states in the particle-number basis. Each of the states forming the basis belongs to one of the classes defined by partitions of integer $N$ into $M$ non-negative integers, i.e. $N = n_1 + n_2 + \ldots + n_M$. However, coherent superpositions between the states belonging to different partitions cannot be locally observed due to the super-selection rules. Only mixtures of such states can be observed. But in any state belonging to a particular partition one of the parties is going to have zero particles. For simpler notation assume that it is the first party, so $n_1 = 0$. Then the state can be factored as $\ket{0} \otimes \ket{\psi}$ for some $\ket{\psi}$. This state is not genuinely multipartite nonlocal. Since the same is true for all states in the mixture the impossibility of observing genuinely multipartite nonlocality when $N<M$ follows.

Whowever, while the condition of one excitation per party is necessary it is not sufficient to guarantee detectable genuinely multipartite nonlocality and/or \emph{useful} entanglement. The mixture resulting from eq. (\ref{eq:Wiseman}) could still contain too many local terms for genuinely multipartite Bell violation to occur. Furthermore, nongenuinely multipartite nonlocality may not attain their maximally allowable violation. To guarantee detectable genuinely multipartite nonlocality one can prepare multiple excitations in the dual rail encoding.

\section{Conclusion}\label{sec:Conclusion}

In this article we have looked at the difference between genuinely and nongenuinely multipartite nonlocality as applied to mode entangled states subject to super-selection rules. We generalized the previous approach from \cite{Heaney10} to adopt the Bell inequalities for multipartite states under SSR. We tested several of the multipartite states and found that they violate nongenuinely multipartite Bell inequalities but not genuinely multipartite ones. We presented the physical and mathematical reasons for why this occurs for our case and proposed that the same conditions apply generally. We find two possible reasons for this. Firstly, the states may not contain enough particles for genuinely multipartite nonlocality. And secondly, the projection of the state to a mixture of states with locally constant number of particles may destroy the correlations. We thus find that the super-selection rules drive the states towards classicality.The advantage of using multiple copies of states is a simple experimental implementation due to creation of two copies of the same state. No reference frames are required.

\bibliographystyle{unsrt}
\bibliography{/home/meznaric/Documents/Entanglement/Entanglement.bib}

\end{document}